\documentstyle[psfig,12pt]{article}

\title{Critical exponents of ferromagnetic Ising model on fractal lattices}

\author{Pai-Yi Hsiao}

\date{\today}

\begin{document}
\maketitle

\begin{center}
\small 
Laboratoire de Physique Th\'eorique de la Mati\`ere Condens\'ee, \\
Universit\'e Paris 7, case 7020, 2 place Jussieu,
75251 Paris Cedex 05, France
\end{center}

\begin{abstract}
We review the value of the critical exponents $\nu^{-1}$,
$\beta/\nu$, and $\gamma/\nu$ of ferromagnetic Ising model on
fractal lattices of Hausdorff dimension between one and three.
They are obtained by Monte Carlo simulation with the help of
Wolff algorithm. The results are accurate enough to show that 
the hyperscaling law $d_f=2\beta/\nu+\gamma/\nu$ is
satisfied in non-integer dimension. Nevertheless, the
discrepancy between the simulation results and the
$\epsilon$-expansion studies suggests that the strong
universality should be adapted for the fractal lattices.
\end{abstract}

\section*{Introduction}
Fractal lattices have self-similar character. Any small part
of them is magnified; the structure similar to the entire one is
observed. They are good candidates to investigate the
physical properties in non-integer dimensions. At
the early of 1980's, Gefen and his coworkers did the pioneer
works on phase transitions in fractal networks 
\cite{gefen80,gefen83,gefen84-1,gefen84-2}. They studied the
ferromagnetic Ising model on fractal lattices by Migdal-Kadanoff
bond-moving real space renormalization group method and found
that the Hausdorff dimension should replace the space dimension
in the description of the critical phenomena. They pointed out
that topological effects play an important role
at criticality. For example, fractals of finite ramification
order can be solved exactly and show no phase transition at finite
temperature; contrarily, fractals of infinite ramification order
exhibit a second order one. Latter, Bhanot et al.
\cite{bhanot84,bhanot85} executed Metropolis Monte Carlo
simulations and suggested the average number of bonds per spin
should be used instead. Bonnier et al.~\cite{bonnier88,bonnier89}
used alternative bond-moving real space renormalization approach
and high temperature expansion as  research tools but
could not close the problems. Few years ago, two
groups~\cite{monceau98,carmona98} performed Monte Carlo
simulations equipped with high efficient Wolff algorithm and
obtained precisely the critical exponents on Sierpinski carpets.
They showed that the hyperscaling law, $d_f = 2\beta/\nu
+\gamma/\nu$, is satisfied for two different fractals of
Hausdorff dimension $d_f$ equal to 1.8927 and 1.7925. Recently,
Hsiao, Monceau, and Perreau~\cite{hsiao00} extended their probes
to the fractals of dimension between 2 and 3, called Sierpinski
sponge, and gave the conclusion reinforcing the previous studies.

\section*{Ferromagnetic Ising model on Sierpinski lattices}
We denote the Sierpinski lattice at $k$th iteration step by the
symbol SP($\ell^d$, $N_{occ}$, $k$) where $\ell$ is the size of
the generating cell, $d$ the dimension of embedding space, and
$N_{occ}$ the number of occupied sites in the generating cell.
The lattice is generated by enlarging the $(k-1)$th iteration one
by replacing each occupied site by the whole generating cell.
Strictly speaking, a Sierpinski lattice is not a true fractal
except  $k$ tends to infinity.  When $d$ is equal
to 2, Sierpinski lattice is called Sierpinski carpet; and when
$d=3$, it is named Sierpinski sponge. In fig.~1 we give two
generators of fractal lattices SP($4^2$, 12) and SP($3^3$, 18),
whose Hausdorff dimension $d_f=\log N_{occ}/\log\ell$ are equal
to 1.792 and 2.631, respectively. We will put a spin on the center
of an occupied site unlike what Gefan et al.~or Bonnier et al.~did
to put a spin on each vertex of an occupied site. We should notice that
putting spins on vertices will cause the ambiguity how to define
the Hausdorff dimension because the total number of spins is not
equal to the total number of occupied sites. The Hamiltonian of
Ising model on a Sierpinski lattice is defined as usual case:
$H(\{s_i\})=-\sum_{<i,j>} s_i s_j$ where $<i,j>$ runs over all
the nearest neighbor bonds on SP($\ell^d$, $N_{occ}$, $k$) and 
the coupling constant has been set to 1. We will
study fractals with infinite ramification order to insure it
exhibits a second order phase transition.

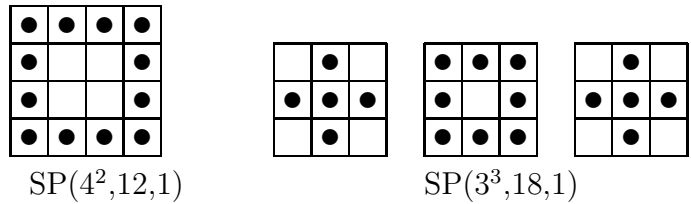
\begin{figure}
\center
\unitlength=0.5cm
\begin{picture}(18,6)
%sp(4^$,12)
\put(0,1){\line(1,0){4}}\put(4,1){\line(0,1){4}}
\put(4,5){\line(-1,0){4}}\put(0,5){\line(0,-1){4}}
\put(0,2){\line(1,0){4}}\put(0,3){\line(1,0){4}}
\put(0,4){\line(1,0){4}}\put(1,1){\line(0,1){4}}
\put(2,1){\line(0,1){4}}\put(3,1){\line(0,1){4}}
\put(0.5,0){SP($4^2$,12,1)}

\put(0.5,1.5){\circle*{0.4}}\put(1.5,1.5){\circle*{0.4}}
\put(2.5,1.5){\circle*{0.4}}\put(3.5,1.5){\circle*{0.4}}
\put(0.5,2.5){\circle*{0.4}}\put(3.5,2.5){\circle*{0.4}}
\put(0.5,3.5){\circle*{0.4}}\put(3.5,3.5){\circle*{0.4}}
\put(0.5,4.5){\circle*{0.4}}\put(1.5,4.5){\circle*{0.4}}
\put(2.5,4.5){\circle*{0.4}}\put(3.5,4.5){\circle*{0.4}}

%sp(3^3,18)
\put(7,1){\line(1,0){3}}\put(10,1){\line(0,1){3}}
\put(10,4){\line(-1,0){3}}\put(7,4){\line(0,-1){3}}
\put(7,2){\line(1,0){3}}\put(7,3){\line(1,0){3}}
\put(8,1){\line(0,1){3}}\put(9,1){\line(0,1){3}}

\put(8.5,1.5){\circle*{0.4}}
\put(7.5,2.5){\circle*{0.4}}\put(8.5,2.5){\circle*{0.4}}\put(9.5,2.5){\circle*{0.4}}
\put(8.5,3.5){\circle*{0.4}}

\put(11,1){\line(1,0){3}}\put(14,1){\line(0,1){3}}
\put(14,4){\line(-1,0){3}}\put(11,4){\line(0,-1){3}}
\put(11,2){\line(1,0){3}}\put(11,3){\line(1,0){3}}
\put(12,1){\line(0,1){3}}\put(13,1){\line(0,1){3}}

\put(11.5,1.5){\circle*{0.4}}\put(12.5,1.5){\circle*{0.4}}
\put(13.5,1.5){\circle*{0.4}}
\put(11.5,2.5){\circle*{0.4}}\put(13.5,2.5){\circle*{0.4}}
\put(11.5,3.5){\circle*{0.4}}\put(12.5,3.5){\circle*{0.4}}
\put(13.5,3.5){\circle*{0.4}}

\put(15,1){\line(1,0){3}}\put(18,1){\line(0,1){3}}
\put(18,4){\line(-1,0){3}}\put(15,4){\line(0,-1){3}}
\put(15,2){\line(1,0){3}}\put(15,3){\line(1,0){3}}
\put(16,1){\line(0,1){3}}\put(17,1){\line(0,1){3}}

\put(16.5,1.5){\circle*{0.4}}
\put(15.5,2.5){\circle*{0.4}}\put(16.5,2.5){\circle*{0.4}}\put(17.5,2.5){\circle*{0.4}}
\put(16.5,3.5){\circle*{0.4}}

\put(11,0){SP($3^3$,18,1)}
\end{picture}
\caption{Generators of Sierpinski lattices: SP($4^2$,12) and SP($3^3$,18).
 SP($3^3$,18,1) embedded in three dimension space is shown section by section
 from left to right. }
\end{figure}

\section*{Homogeneity hypothesis and finite size scaling}
The generalized homogeneity of the free energy per spin in
integer space dimension $d$ was first proposed by Widom
\cite{widom65} and later by Wilson renormalisation group approach
\cite{wilson69}. It states that near the critical point, the free
energy per spin behaves as $ f(t,h,L) = b^{-d}
f(tb^{yt},hb^{yh},L/b) $ under the change of length scale from
$1$ to $b$, where $t$ is the reduced temperature $(T-T_c)/T_c$,
$h$ the external magnetic field, and $L$ the size of the system.
Due to the scale invariant character of fractal lattices, one can
extend the above hypothesis to fractals by directly replacing the
space dimension $d$ by the Hausdorff one $d_f$. The immediate
consequence is that the critical exponents should satisfy the
following two relations: $ \alpha + 2\beta +\gamma =2 , \ \
d_f\nu=2\beta +\gamma$. According to Fisher's finite size scaling
theory \cite{fisher72}, for a lattice of finite size $L$, the
specific heat per spin $C_L(t)$, the thermal average of the
absolute value of magenetization per spin $m_L(t)$, the magnetic
susceptibility per spin $\chi_L(t)$, and the logarithmic
derivative of magenetization $\phi_L(t)= \partial \ln m_L(t) /
\partial \beta_B$ should behave like following:
\begin{equation}
\Omega_L(t)=L^{\Delta_{\Omega}} {\cal Q}_{\Omega}(tL^{1/\nu})
\end{equation}
where $\Omega$ stands for $C$, $m$, $\chi$, and $\phi$ and
$\Delta_{\Omega}$ equal to $\alpha/\nu$, $-\beta/\nu$,
$\gamma/\nu$, and $\nu^{-1}$, respectively. We can see that the
maximum value of physical quantity $\Omega$ is determined by the
function ${\cal Q}_{\Omega}(x)$ with $x=tL^{1/\nu}$. The critical
exponents $\alpha/\nu$, $\gamma/\nu$, and $\nu^{-1}$ can be,
hence, extracted by finding the associated maximum values
$\Omega_L^{max}$ at different lattices sizes. Moreover, if 
the maximum occurs at $T_{\Omega}(L)$ for a system of size $L$,
the critical temperature $T_c$ for a lattice of infinite size can
be extrapolated with the help of the equation
$T_{\Omega}(L)=T_c(1+x_{\Omega}^* L^{-1/\nu})$. Therefore we can
execute Monte Carlo simulations at criticality $T_c$ and extract
the exponents $\alpha/\nu$, $\beta/\nu$, $\gamma/\nu$, and
$\nu^{-1}$ directly.

\section*{Monte Carlo simulation}
We setup periodic boundary condition on the studied lattices and
perform Monte Carlo simulation. Wolff algorithm
\cite{wolff88} is applied at some temperature for producing spin
configurations.  Histogram method is then used to extract all possible
 information near this temperature.  We
repeat the whole above processes several times to get a
sufficient number of samples for doing statistical analysis.

\section*{Results and discussions}
In table 1, we list the critical temperatures and the critical
exponents $\nu^{-1}$, $\beta/\nu$, and $\gamma/\nu$ of the five
different fractal lattices. The results of SP($4^2$,12) and
SP($3^2$,8) are taken from the paper of Carmona et al.~\cite{carmona98} 
in which the two lattices are denoted by
fractal B and fractal A, respectively.  The three fractals of 
dimension between 2 and 3 are reported in our recent article
\cite{hsiao00}. The effective dimensions
$2\beta/\nu+\gamma/\nu$ of the five fractals can be calculated to
be 1.808(40), 1.890(3), 2.640(18), 2.900(23), and 2.955(41),
respectively; they are consistent with their Hausdorff dimensions.
It proves the validity of homogeneity hypothesis for fractals.

\begin{table}[t]
\center
\begin{tabular}{ |c|ccccc|}
 \hline
SP & $(4^2,12)^a$ & $(3^2,8)^a$ & $(3^3,18)^b$ & $(4^3,56)^b$ & $(3^3,26)^b$ \\
  \hline
  $d_f$  & 1.792 & 1.893 & 2.631 & 2.904 & 2.966 \\
  $T_c$  &1.078(3) & 1.4813(2) & 2.35090(9) & 3.99893(10) & 4.21701(6) \\
  $1/\nu$&0.309(8) & 0.59(1) & 1.185(27) & 1.410(36) & 1.503(53) \\
  $\beta/\nu$& 0.069(10) & 0.080(1) & 0.3224(14) & 0.493(5) & 0.506(11) \\
  $\gamma/\nu$& 1.67(2) & 1.730(1) & 1.995(15) & 1.914 (13) & 1.943(19)\\
  \hline
\end{tabular}
\caption{The critical temperatures and exponents of five
different fractals. $^a$ Data reproduced from the article of Carmona et al. 
[10].  $^b$ Results reported in [11]. }
\end{table}

The effect of correction-to-scaling can be observed in the studies. 
It is found to be weaker in the case of dimension between 2 and 3  
than that between 1 and 2. This can be understood by considering the
convergence speed of the deviation
ratio $\rho(\ell^d, N_{occ}, k)$ which is defined as 
the percentage of the difference between the mean number of neighbors
at step $k$ and that at infinite step.
$\rho(\ell^d,N_{occ},k)$ measures the discrepancy between the 
{\it pseudo}-fractal and the true one and can be shown to be:
\begin{equation}
\left[\frac{(N_{occ}- N_S)d}{N_I} -1\right] \left( \frac{N_S}{N_{occ}}\right)^k
\end{equation}
where $N_S$ is the number of occupied sites on each surface (boundary) 
of the generating cell and $N_I$ the number of its internal bonds.
As $k$ increases, $\rho(\ell^d,N_{occ},k)$ tends to zero faster than the
fractals embedded in a two dimensionnal space.

Physicists believe that the critical properties do not depend on the structure
of the lattice but only on the dimension of system, the range of couplings, 
and the symmetry of order parameter. This is  called the strong universality.
According to it, the critical exponents of a fractal lattice should fall in the
same class like the hypercubic one of the same dimension. 
But comparing with the values calculated by $\epsilon$-expansion 
\cite{guillou87}, we find that they are not consistent with each other. 
The simulation results reinforce the argument of Gefen et al.~that the 
topology effects of a fractal play an indispensable role in determining 
its critical properties.  
Strong universality should be revised for self-similar lattices.

\section*{Acknowledgments}
The author acknowledges the financial support from the ICSC World
Laboratory in Switzerland.

\end{document}